\documentclass[sigconf]{acmart}

\AtBeginDocument{%
  }
\usepackage[table]{xcolor}
\copyrightyear{2026}
\acmYear{2026}
\setcopyright{cc}
\setcctype{by-nc-nd}
\acmConference[WWW '26]{Proceedings of the ACM Web Conference 2026}{April 13--17, 2026}{Dubai, United Arab Emirates}
\acmBooktitle{Proceedings of the ACM Web Conference 2026 (WWW '26), April 13--17, 2026, Dubai, United Arab Emirates}
\acmPrice{}
\acmDOI{10.1145/3774904.3792563}
\acmISBN{979-8-4007-2307-0/2026/04}

\usepackage{graphicx}
\usepackage{amsmath, amsthm, amsfonts}
\usepackage{dsfont}
\usepackage{wrapfig}
\usepackage{algorithm} 
\usepackage{algorithmic}
\usepackage{multirow}
\usepackage{makecell}

\usepackage[utf8]{inputenc} 
\usepackage[T1]{fontenc}    
\usepackage{hyperref}       
\usepackage{url}            
\usepackage{booktabs}       
\usepackage{amsfonts}       
\usepackage{nicefrac}       
\usepackage{microtype}      
\usepackage{xcolor}         
\usepackage{enumitem}
\newtheorem{theorem}{Theorem}[section]
\newtheorem{proposition}[theorem]{Proposition}

\definecolor{rowA}{RGB}{255,220,230} 
\definecolor{rowB}{RGB}{220,235,255} 
\definecolor{col1}{RGB}{255,245,220} 
\definecolor{col2}{RGB}{220,255,230} 

\definecolor{blend11}{RGB}{255,233,225} 
\definecolor{blend12}{RGB}{238,240,230} 
\definecolor{blend21}{RGB}{238,238,240} 
\definecolor{blend22}{RGB}{220,245,240} 
\begin{document}
\title{Quantifying User Coherence: A Unified Framework for Analyzing Recommender Systems Across Domains}

\author{Michaël Soumm}
\authornote{Both authors contributed equally to this research.}
\email{soumm@telecom-paris.fr}
\orcid{1234-5678-9012}
\affiliation{%
  \institution{Télécom Paris}
  \city{Palaiseau}
  \country{France}
}

\author{Alexandre Fournier Montgieux}
\authornotemark[1]
\email{alexandre.fourniermontgieux@cea.fr}
\affiliation{%
  \institution{CEA List}
  \city{Palaiseau}
  \country{France}
}

\author{Adrian Popescu}
\email{adrian.popescu@cea.fr}
\affiliation{%
  \institution{CEA List}
  \city{Palaiseau}
  \country{France}
  }
\author{Bertrand Delezoide}
\email{bertrand.delezoide@amanda.com}
\affiliation{%
  \institution{Amanda.com}
  \city{Paris}
  \country{France}
}

\renewcommand{\shortauthors}{Michaël Soumm, Alexandre Fournier Montgieux, Adrian Popescu, \& Bertrand Delezoide}

\begin{abstract}

The performance of Recommender Systems (RS) varies significantly across users, yet the underlying reasons for this variance remain poorly understood. This paper introduces a unified framework to analyze and explain this performance gap by quantifying user profile characteristics. We propose two novel, information-theoretic measures: \textbf{Mean Surprise ($S(u)$)}, which captures a user's deviation from popular items and is closely related to popularity bias, and \textbf{Mean Conditional Surprise ($CS(u)$)}, which measures the internal coherence of a user's interactions in a domain-agnostic manner. Through extensive experiments on 7 algorithms and 9 datasets, we demonstrate that these measures are strong predictors of recommendation performance. Our analysis reveals that performance gains from complex models are concentrated on "coherent" users, while all algorithms perform poorly on "incoherent" users. We show how these measures provide practical utility for the Web community by: (1) enabling robust, stratified evaluation to identify model weaknesses; (2) facilitating a novel analysis of the behavioral alignment of recommendations; and (3) guiding targeted system design, which we validate by training a specialized model on a segment of "coherent" users that achieves superior performance for that group with significantly less data. This work provides a new lens for understanding user behavior and offers practical tools for building more robust and efficient large-scale recommender systems.
\end{abstract}

\ccsdesc[500]{Information systems~Personalization}
\ccsdesc[500]{Information systems~Recommender systems}

\keywords{Recommender Systems, Explainability, User profile quality, Metrics, Algorithm performance}
\maketitle

\section{Introduction}

With the rapid development of digital connectivity and the explosion in the amount of information, recommender systems (RS) have emerged as essential tools, offering personalized suggestions adapted to individual preferences and behaviors \cite{Nilashi2013CollaborativeFR, recommender_pavlidis_2019}. 
These systems play an essential role in filtering and personalizing content across various domains, from e-commerce to entertainment and news consumption \cite{Nilashi2013CollaborativeFR, introduction_konstan_2004}.

\begin{figure}[t]
  \centering
    \includegraphics[width=\linewidth]{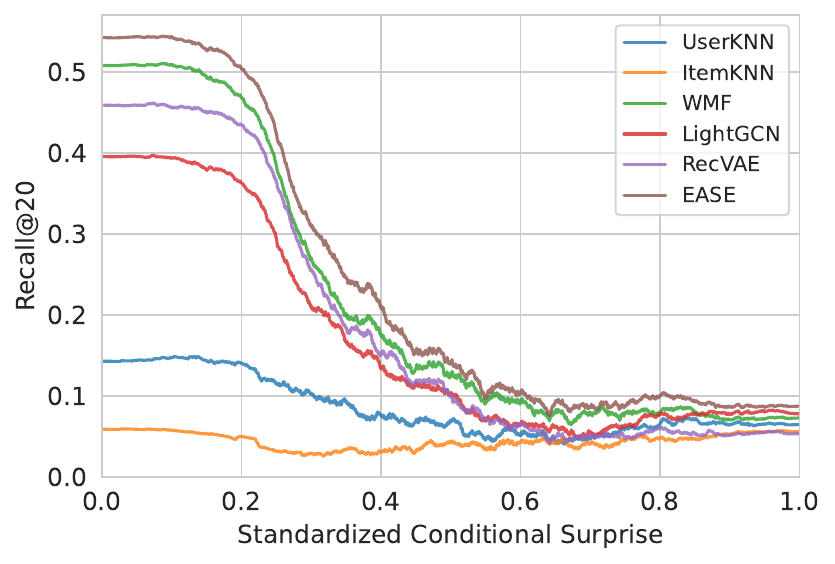}
  \caption{Performance in \texttt{Recall@20} of different RS, averaged over datasets, w.r. to our proposed \textbf{Conditional Surprise} ($CS(u)$) measure, which represents \textbf{how much a user's interactions are coherent}, standardized between 0 and 1. All RS performance collapse for high values of $CS(u)$, showing that most model gains stem from learning only on "easy" data points.}
  \label{fig:side-by-side}
  \vspace{-10pt}
\end{figure}

RS can be categorized into three main types: content-based (CB),  knowledge-based (KB), and in our study scope, collaborative filtering (CF) approaches \cite{knowledgebased_burke_2000}. We concentrate on CF approaches since they have dominated the research landscape in recent years \cite{trends_lops_2019, zhang2019deep, batmaz2019review}: Their ability to aggregate user preferences and make recommendations based on similarities in user behavior patterns \cite{introduction_konstan_2004} despite minimal information allows analyzing only user-item interactions without the need for additional content or knowledge-based features \cite{collaborative_ekstrand_2011, collaborative_afoudi_2018}.

In the context of deep learning, inconsistency within data appears to be a central issue for both training and model evaluation \cite{classnoise}. Where some applications seek to eliminate these parasitic components \cite{dealingwithnoise, Learning_deep_attribution_priors}, others seek to better deal with them, or even propose evaluation frameworks dedicated to these more “difficult” scenarios \cite{gale2025statisticalcoherencealignmentlarge, favcid}. 
However, in both cases, introducing effective metrics or measures to analyze and characterize the data is essential.  In particular, selecting the right measures is required for ensuring that an RS delivers recommendations that are not only accurate but also align well with users' overall preferences and consumption patterns, ultimately enhancing user satisfaction. 
Despite the emergence of many new measures like diversity and novelty \cite{diversity_kaminskas_2016} that describe predicted items, and despite existing research on users' profiles classification \cite{combining_claypool_1999, fulfilling_ghazanfar_2011}, the field lacks measures that accurately model the diversity of user consumption.

We thus propose two coherence measures, \textbf{Mean Surprise} and \textbf{Mean Conditional Surprise}, that quantify how “unusual” user profiles are. The former describes how surprising (uncommon) a user's consumption is, while the latter describes the consumption's internal coherence, independently of its uncommonness. Both measures are strongly linked with the models' prediction quality, especially the conditional surprise as summarized in Figure \ref{fig:side-by-side}. Specifically, our main contributions are:
\begin{itemize}[noitemsep, topsep=0pt, partopsep=0pt, parsep=0pt, leftmargin=0.3cm]
    \item We introduce a framework based on two robust, information-theoretic measures, \textbf{Mean Surprise $S(u)$} and \textbf{Mean Conditional Surprise $CS(u)$}, to quantify user taste and profile coherence across diverse domains.
    
    \item We use these measures to perform a \textbf{stratified evaluation} of recommender systems, revealing a key insight that aggregate metrics hide: performance gains from complex models are concentrated on "coherent" users, while all algorithms universally fail on "incoherent" users.
    \item We demonstrate the practical utility of our framework for deeper algorithmic analysis and system design by: (a) proposing a novel method to measure the \textbf{behavioral alignment} of algorithms, and (b) validating \textbf{user segmentation} with a proof-of-concept where a specialized model improves performance on coherent users with less data.
\end{itemize}
The anonymized version of the code can be found at \url{https://github.com/MSoumm/coherence-measures}.

\begin{table*}[t]
\centering
\caption{User profile examples of the archetypes defined by our proposed measures.}
\label{tab:archetypes}
\resizebox{\linewidth}{!}{%
\begin{tabular}{c||c|c}

& \textbf{Low $CS(u)$} (Coherent) &  \textbf{High $CS(u)$} (Incoherent) \\
\midrule
 \textbf{Low $S(u)$} (Mainstream) &   \cellcolor{col2}Popular, related series (e.g., \textit{The Avengers}, \textit{Harry Potter}).& \cellcolor{rowB} Popular but unrelated blockbusters (e.g., \textit{Frozen}, \textit{Joker}, \textit{Avatar}). \\
\hline
\textbf{High $S(u)$} (Niche) & \cellcolor{col1} A single unknown director's filmography. &  \cellcolor{rowA} An eclectic mix of rare, unrelated films (e.g., documentary, horror, Bollywood). \\

\end{tabular}%
}
\end{table*}
\section{Related Work}
The evaluation of Recommender Systems (RS) has matured significantly, with a growing emphasis on reproducibility and rigor \cite{evaluating_wu_2012, systematic_roy_2022}. This has been driven by challenges in offline evaluation \cite{challenges_sun_2023, beyond_sun_2024} and replicability \cite{newer_dong_2023}, motivating the need for more rigorous evaluation approaches.

The data processing pipeline is also known to have an important impact on performance. Some works \cite{exploring_meng_2020} examined data splitting strategies and their impact on evaluation outcomes, while others \cite{critical_ji_2020} addressed data leakage in offline evaluation, highlighting the issue of evaluating on too few sets \cite{model_fan_2024}.
These concerns lead to the development of standardized frameworks like Elliot \cite{elliot_anelli_2021} and ReChorus2.0 \cite{rechorus20_li_2024}. While essential, these frameworks have primarily centered on traditional accuracy metrics.

Recognizing the limitations of accuracy-only evaluation, the community developed metrics that assess the quality of the \textit{recommendation list} (the system's output). These include measures of diversity, novelty, serendipity, and user satisfaction \cite{diversity_kaminskas_2016, recommender_konstan_2012, good_silveira_2019}. The use of satisfaction metrics was democratized to evaluate model predictions in the following works \cite{good_silveira_2019, survey_alhijawi_2022}. Additional research has highlighted the importance of considering these newer metrics in order to better match users' behavior and find better recommendations \cite{customer_kim_2021, beyond_ping_2024}. Such metrics are useful for analyzing system behavior and group disparities \cite{newer_dong_2023, beyondaccuracy_urii_2023}. However, they primarily evaluate the outcome of a recommendation without fully explaining \textit{why} performance varies across users. Some model-based surprise approaches, such as Unexp-DIN\cite{UnexpDIN}, focus on unexpectedness relative to a specific architecture's predictions, our measures are model-agnostic. They characterize the intrinsic properties of the input data itself, serving as a complementary tool to analyze why certain architectures may fail to capture specific user behavior patterns

Explaining this variance requires characterizing the \textit{input} data: the user profiles themselves. A notable line of work is the identification of "grey sheep" users, whose preferences are hard to model as they do not align with any discernible group \cite{combining_claypool_1999}. Existing methods for identifying such users often rely on inter-user similarity clustering or density-based techniques \cite{fulfilling_ghazanfar_2011, identification_zheng_2017, IdentifyingGreySheep}. While insightful, these approaches can be sensitive to profile density and often require dataset-specific hyperparameter tuning, limiting their generalizability.

In addition, concerns about domain-specific evaluation methodologies have emerged, highlighting the need for more nuanced and versatile evaluation approaches.   

 Recent findings \cite{streaming_latifi_2022, understanding_dietz_2023} underscore the importance of considering domain-specific features in evaluation methodologies. Building on these insights, \citet{beyond_sun_2024} questioned the applicability of current evaluation practices across domains, arguing that metrics and evaluation protocols optimized for one domain may not translate effectively to others. They highlighted the need for more generalizable evaluation frameworks that can account for domain-specific nuances while still enabling meaningful cross-domain comparisons.

Collectively, these studies depict a field in transition, showing the need for better evaluation practices to match the increasing sophistication and real-world impact of recommender systems.

\section{Proposed Approach} 
\subsection{Notations and Problem Formulation for Binary Collaborative Filtering}
We denote the set of users as \(\mathcal{U}\) and the set of items as \(\mathcal{I}\). Let $n= |\mathcal{U}|$ be the number of users and $m=|\mathcal{I}|$ the number of items. We place ourselves in a binary setting where each user \(u \in \mathcal{U}\) can either interact (\(x_{ui} = 1\)) or not interact (\(x_{ui} = 0\)) with an item \(i \in \mathcal{I}\), leading to the binary interaction matrix $X \in \{0,1\}^{n\times m}$. We identify each user to its item set so that "$u$" refers equivalently to a user id and to their item set. \\
A set of test users \(\mathcal{U}_{test}\) is randomly sampled. For each test user, their last interaction is isolated as the test target $x_{ui_{test}}$. All other interactions are part of the training set. The goal of a recommender system is to learn a function $f: \mathcal{U} \times \mathcal{I} \rightarrow \mathbb{R}$ that assigns a score $f(u,i)$ to each user-item pair $(u,i)$, indicating the likelihood of user $u$ interacting with item $i$. The system's performance is evaluated by its ability to rank the test item $i_{test}$ highly among all items not in the user's training set.

\subsection{Coherence measures}
We aim to understand and quantify individual user behavior in recommender systems. Traditional approaches often simplify users into entries in a user-item matrix, and this modeling choice can overlook important nuances in user preferences and consumption patterns. By modeling individual user behavior more comprehensively, we can gain insights into why certain recommendations succeed or fail, and potentially tailor our approaches to different types of users.

In this context, we define coherence as "\textit{the degree to which a user's interactions form a consistent and predictable pattern}". A highly coherent user would have a set of interactions that align well with each other and with common consumption patterns. In contrast, a less coherent user might have more diverse interactions.

Specifically, we want to measure how the performance of recommender algorithms is impacted by how surprising or unpredictable a user's behavior is. To model surprise, one natural way is to first assign probabilities to items. In previous works \cite{diversity_kaminskas_2016}, the probabilities used to compute existing measures describe the prediction distribution. However, this approach is limited as it focuses on the model's output rather than the inherent characteristics of user behavior. Here, we adopt another view and consider the item probability as their frequency among the users:
\begin{equation}
    p_i^* =\frac{\big|\{x_{ui} = 1, u\in \mathcal{U}\}\big|}{n} 
\end{equation}
This quantity is insufficient since a user can interact with very rare items but still have a very coherent set of items (e.g. niche movies; but all from the same director). Therefore, we consider the second-order statistics:
\begin{equation}
    p_{i,j}^* =\frac{ \big|\{x_{ui} = x_{uj} =1, u \in \mathcal{U}\}\big|}{n}  \quad \textrm{and} \quad p_{i|j}^* =\frac{p_{i,j}^*}{p_j^*}
\end{equation}
The probability $p_{i|j}^*$ represents how much a user is likely to interact with $i$ when they also interacted with $j$.\\
\indent Then, we can define the \textbf{Surprise} (or novelty, as in \cite{castells2011novelty}) of an item as $ -\log(p_i^*)$ and the \textbf{Conditional Surprise} $ -\log (p_{i|j}^*)$. The first intuitive measures to study are the mean empirical binary cross-entropies:

\begin{align}
    \label{eq:vanilla_ce}
    \widetilde{S}(u) &= -\frac{1}{m}\sum_{i=1}^m \log(p_i^*)x_{ui} &
    \widetilde{CS}(u) &= -\frac{1}{m^2}\sum_{i=1}^m\sum_{j=1}^m \log(p_{i|j}^*)x_{ui}x_{uj}
\end{align}

However, these existing definitions have an important limitation in that they are non-decreasing when a new item is added to a user's set, regardless of the item's characteristics. This behavior is counter-intuitive, as we would expect the overall surprise to potentially decrease if a highly predictable or common item is added to the user's profile. Thus, we define our user coherence measures, called \textbf{Mean Surprise} and \textbf{Mean Conditional Surprise}:
\begin{align}
\label{eq:ours_ce}
    S(u) &= -\frac{1}{|u|}\sum_{i\in u}\log(p_i^*) &
    CS(u) &= -\frac{1}{|u|^2}\sum_{i\in u}\sum_{j\in u}\log(p_{i|j}^*)
\end{align}
where $|\cdot|$ is the $L^1$ norm.
To show the relevance of our measures, we also compute their Oracle versions on the test items. In our leave-one-out setup, they simplify to $S(i_{test}) = -\log(p_{i_{test}})$ and $CS(i_{test}|u) = -\frac{1}{ |u|}\sum_{j\in u}\log(p_{i_{test}|j})$.

Contrary to classical novelty definitions \cite{castells2011novelty,silveira2019good} that are applied to the recommended items, our measures focus on the already consumed items to model user behavior.
\paragraph{Qualitative Examples.} To build intuition, our measures define four main user archetypes based on their taste (mainstream vs. niche) and coherence (consistent vs. incoherent). The Table \ref{tab:archetypes} illustrates these archetypes for a movie domain.

\paragraph{Interpretation and Properties.} The quantities in Equations \ref{eq:ours_ce} have a similar form as in Equations \ref{eq:vanilla_ce}, but provide a dynamic rescaling dependent on the user. This ensures that the measures are comparable across users with different numbers of interactions, providing a fair basis for comparison. Unlike the previous formulations, these measures can decrease when a user interacts with a common item, better reflecting intuitive notions of surprise and coherence. The Mean Surprise $S(u)$ describes at the first order how much a user's consumption deviates from the popular items, on a scale from the \textit{unsurprising} users to the \textit{surprising} users. The Mean Conditional Surprise $CS(u)$ indicates whether the co-occurrences in the user's consumption set are far from frequent co-occurrences, capturing the internal consistency of a user's choices, on a scale from the \textit{coherent} users to the \textit{incoherent} users.

In RS data, user behavior and item consumption detection can be quite noisy \cite{reco_noise}. We can verify how our measures behave on average:
\begin{proposition}
Let $\pi_u$ be the distribution from which $u$ is drawn, and  $\pi_u^{\geq 1}$ be the distribution of $u$ conditioned on $|u| \geq1$. Let $S^*(u)=\mathbb{E}_{\pi_u^{\geq 1}}[\widetilde{S}(u)]$ and $CS^*(u)= \mathbb{E}_{\pi_u^{\geq 1}}[\widetilde{CS}(u)]$. Then (Proof in \ref{sup:proof1}):

\begin{align}
    \frac{m}{\mathbb{E}_{\pi_u^{\geq1}}[|u|]}&\leq \frac{\mathbb{E}_{\pi_u^{\geq1}}[S(u)]}{S^*(u)}\leq \mathbb{E}_{\pi_u^{\geq1}}\left[\frac{m}{|u|}\right]
    \\
    \frac{m^2}{\mathbb{E}_{\pi_u^{\geq1}}[|u|^2]}&\leq \frac{\mathbb{E}_{\pi_u^{\geq1}}[CS(u)]}{CS^*(u)}\leq \mathbb{E}_{\pi_u^{\geq1}}\left[\frac{m^2}{|u|^2}\right]
\end{align}

\end{proposition}
In particular, we see that the lower bound for the scaling depends only on the expected value of the number of items. The upper bound, however, can theoretically get bigger than $m/\mathbb{E}_{\pi_u^{\geq1}}[|u|]$. Since $|u|$ is the number of items consumed by $u$, we can model it as a Poisson variable with parameter $\lambda$ (\ref{sup:PoissonModel}). We have the following bound (Proof in \ref{sup:prop2}):
\begin{proposition}
    If $X$ is a Poisson variable of parameter $\lambda>0$ , we have:
    \begin{equation}
    \label{eq:poisson}
          \mathbb{E}_{\geq1}\left[\frac{1}{X}\right]\leq \frac{2}{\mathbb{E}_{\geq1}[X]}
    \end{equation}
\end{proposition}
Empirically, we find a tighter upper bound for Equation \ref{eq:poisson} with a numerator equal to 1.37 instead of 2. This upper bound is met for $\lambda_{sup}\approx2.9$. For smaller and larger values of $\lambda$, the numerator quickly drops to 1. This means that for surprising users, on average, the empirical estimation should not be too far from $S^*(u)\times m/{\mathbb{E}_{\pi_u^{\geq1}}[|u|]}$, which is indeed the classical estimator expectancy re-scaled by the mean proportion of items that a user will consume. The same bound effects apply to $CS(u)$.

\paragraph{Implications of Theoretical Guarantees.}
These propositions provide the main theoretical link between our practical, user-level measures and the ideal cross-entropy we aim to approximate ($S^*(u)$ and $CS^*(u)$). The core challenge is that the naive estimators are poorly suited for sparse binary data. Our proposed measures resolve this by normalizing only over a user's consumed items. Propositions 1 and 2 together demonstrate the consequence of this choice: on average, our proposed estimator is a stable, well-behaved rescaling of the ideal cross-entropy we wish to target. The bounds show that this rescaling factor is tightly controlled by the inverse of the user base's average consumption rate (proportional to $m/\mathbb{E}[|u|]$). This result provides confidence that our measures are not ad-hoc heuristics; they are principled estimators that effectively magnify the relevant behavioral signal from sparse interactions in a theoretically justified manner.

\subsection{Regression Model as an Analytical tool}
\label{sec:regression_models_explained}
To accurately quantify the impact of our measures on RS performance, we employ logistic regression \cite{mostlyharmless}, a statistical method for modeling the relation between attributes and binary outcomes, as proposed by \citet{soumm2024causalinferencetoolsbetter}. The logistic regression model is defined as:
\begin{equation}
\ln \dfrac{\mathbb{P}[y=1|X]}{\mathbb{P}[y=0|X]} = \beta_0 + \beta_1 X_1 + ... +\beta_k X_k
\end{equation}
where $y$ is the binary target variable, $X_1, ...X_k$ are the explanatory variables, and $\beta_0, ..., \beta_k$ are the fitted coefficients. This model is denoted in a condensed form as $Y \sim \sigma(X_1 + ... + X_k)$.
The model provides coefficient estimates ($\hat{\beta}$ values) for each variable, from which we compute average marginal effects (AME), quantifying how changes in each variable affect the outcome probability while holding other variables constant \cite{mostlyharmless}. The choice of the considered variables is motivated by modeling considerations and regression quality metrics such as McFadden $R^2$ \cite{allison_measures_nodate} and the AIC \cite{akaike1998information}.
To capture the interactions between variables, we use product terms. For example, the notation $X \times Y$  expands to $X + Y + XY$, including the main effects and their interaction. This allows for the modeling of complex relationships while retaining individual variable effects.

When $X$ is a variable estimating $X^*$ with a certain variance $\sigma^2$, the plain regression on $X$ becomes imprecise, as the regression data becomes noisy. We the SIMEX (Simulation-Extrapolation) method \cite{cook1994simulation}, which simulates many regressions with the added noise $\sigma^2$, and extrapolates to the case of no noise, providing more robust coefficient estimates.

\section{Experimental Setup}
\begin{table}[t]
\caption{Description of the datasets after processing (see \ref{sec:dataprocessing} for details about the processing)}
    \label{tab:datasets}
    \centering
     
   \begin{tabular}{@{}c||c|c|c|c}
   \toprule
 Dataset & Items & Users & Inter. & Density\\
\midrule
ML 1M & 3.1K & 6K & 562K  & $3\cdot10^{-2}$\\
ML 10M & 9.4K & 69K & 5.7M  & $9\cdot10^{-3}$\\
Netflix Small & 2.7K & 8.3K & 320K  & $1\cdot10^{-2}$\\
Netflix & 18K & 463K & 59.9M & $7\cdot10^{-4}$\\
Vis2Rec & 9.3K & 9.1K & 200K  & $2\cdot10^{-3}$\\
Tradesy & 12K & 6.6K & 73K  & $9\cdot10^{-4}$\\
Amazon Music & 11K & 8.6K & 87K  & $9\cdot10^{-4}$\\
Amazon Office & 62K & 20K & 468K  & $4\cdot10^{-4}$\\
Amazon Toys & 143K & 61K & 1.2M  & $1\cdot10^{-4}$\\
\bottomrule
\end{tabular}
 \vspace{-0.45cm}
\end{table}
\subsection{Datasets}
We perform our analysis on 9 datasets of various sizes and domains:
\textbf{MovieLens 1M} and {\textbf{MovieLens 10M}} \cite{movielens_harper_2016}: movie ratings recommendation datasets, commonly used for benchmarking recommendation algorithms;
 {\textbf{Netflix Small}}  and {\textbf{Netflix}} \cite{BeLa07}: two versions of the Netflix Prize dataset, consisting of movie ratings;
 {\textbf{Amazon Music}}, {\textbf{Amazon Office}}, and {\textbf{Amazon Toys}} 
\cite{name_lakkaraju_2013}: part of the Amazon product reviews collection, focusing on different product categories;
 {\textbf{Tradesy}} 
\cite{name_lakkaraju_2013}: interactions on the Tradesy platform, which specializes in the resale of designer fashion;
 {\textbf{Vis2Rec}} \cite{vis2rec_soumm_2023}: tourism recommendation dataset, with visits identified from photos.

We used different sizes from the same dataset, and datasets from the same domain but different sources, specifically to study the effects of dataset size and domain. This allows us to analyze recommender systems across a wide range of real-world applications.

\begin{figure*}[t]
    \centering
\includegraphics[width=0.95\linewidth]{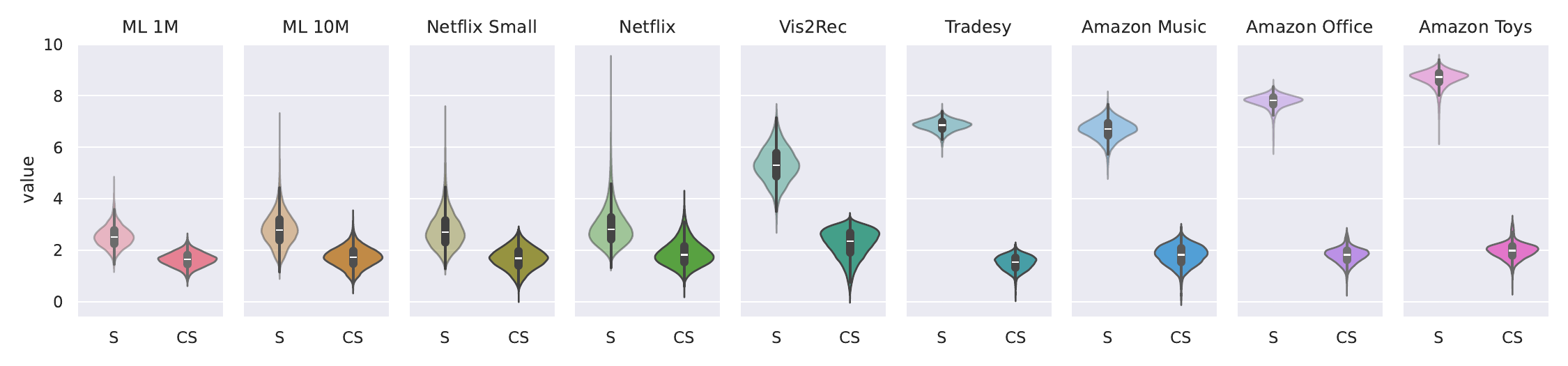}
    \caption{Distribution of the measures across datasets. S denotes the Surprise measure, and {CS} the Conditional Surprise measure. CS shows remarkable stability across all datasets.}
    \label{fig:distribution}
    \vspace{-10pt}
\end{figure*}
\subsection{Recommender Algorithms}

We benchmark 7 recommendation algorithms on the formated datasets presented in Table \ref{tab:datasets}  :
\texttt{\textbf{MostPop}} is the baseline algorithm that recommends the most popular items to every user;
\texttt{\textbf{UserKNN}} is a neighborhood-based approach that relies on a similarity between pairs of users.
\texttt{\textbf{ItemKNN}}  is a neighborhood-based approach that relies on a similarity between pairs of items;
\texttt{\textbf{WMF}} is a weighted matrix factorization approach that learns user and item embeddings with gradient descent, minimizing a reconstruction loss;
\texttt{\textbf{EASE}} \cite{Steck_2019} is a matrix factorization approach that computes an item-item weight matrix with a closed-form formula;
\texttt{\textbf{LightGCN}} \cite{Shenbin_2020} is an approach that learns user and item embeddings by aggregating information from the user-item interaction graph using a light graph convolution;
\texttt{\textbf{RecVAE}} \cite{Shenbin_2020} is a variational auto-encoder approach inspired by $\beta$-VAE \cite{higgins2017betavae} and denoising-VAE \cite{im2016denoisingcriterionvariationalautoencoding}.
These algorithms have been chosen to represent a wide range of possible usages, depending on how well they scale with the number of users, the number of items, training times, or flexibility.
\subsection{Training}
\label{sec:training_subsec}
In the leave-one-out protocol, as there is only 1 relevant test item per user, the two reference metrics \texttt{Recall@K} and \texttt{Precision@K} \cite{herlocker2004evaluating} are equivalent, up to a constant, since they are proportional to the number of relevant items. Therefore, we only use  \texttt{Recall@K}, which corresponds in this case to a binary variable 0/1.
For each experiment (i.e. algorithm trained on a dataset or a dataset segment), we perform a hyperparameter search using \texttt{optuna} with 50 rounds maximizing the \texttt{Recall@20} on the validation set.The models are trained on 256 AMD EPYC 9554 64-Core CPUs and 1.4TB of RAM for the algorithms that run on CPU, while others are run on a single NVIDIA A100 GPU with 40GB of VRAM.

\section{Results and Analysis}

\subsection{Experimental Properties of the Measures}
\paragraph{Measure Distribution.} Figure \ref{fig:distribution} presents the distribution of our proposed measures across the datasets. A key observation is that the distribution of $S(u)$ characterizes the domain: the movie-related datasets exhibit comparable Mean Surprise values. This suggests a uniformity in user behavior patterns within the movie recommendation domain, even if they come from different sources and collection processes.
In contrast, all e-commerce datasets demonstrate higher $S(u)$ values, indicating higher diversity in user consumptions. Vis2Rec, designed for tourism recommendation,  falls between the movie and e-commerce clusters, highlighting its distinct nature.

Interestingly, despite the variations in the distribution of $S(u)$ across domains, we observe consistency in the distribution of $CS(u)$ across all datasets. This suggests that the $CS(u)$ measure is a good candidate for a domain-agnostic coherence measure.

\paragraph{Comparison with naive measures.} We start by verifying that our measures behave better than existing ones, such as the mean-cross entropies $\widetilde{S}(u)$ and $\widetilde{CS}(u)$ defined in Eq. (\ref{eq:vanilla_ce}).

\begin{figure}
  \centering
  \includegraphics[width=\linewidth]{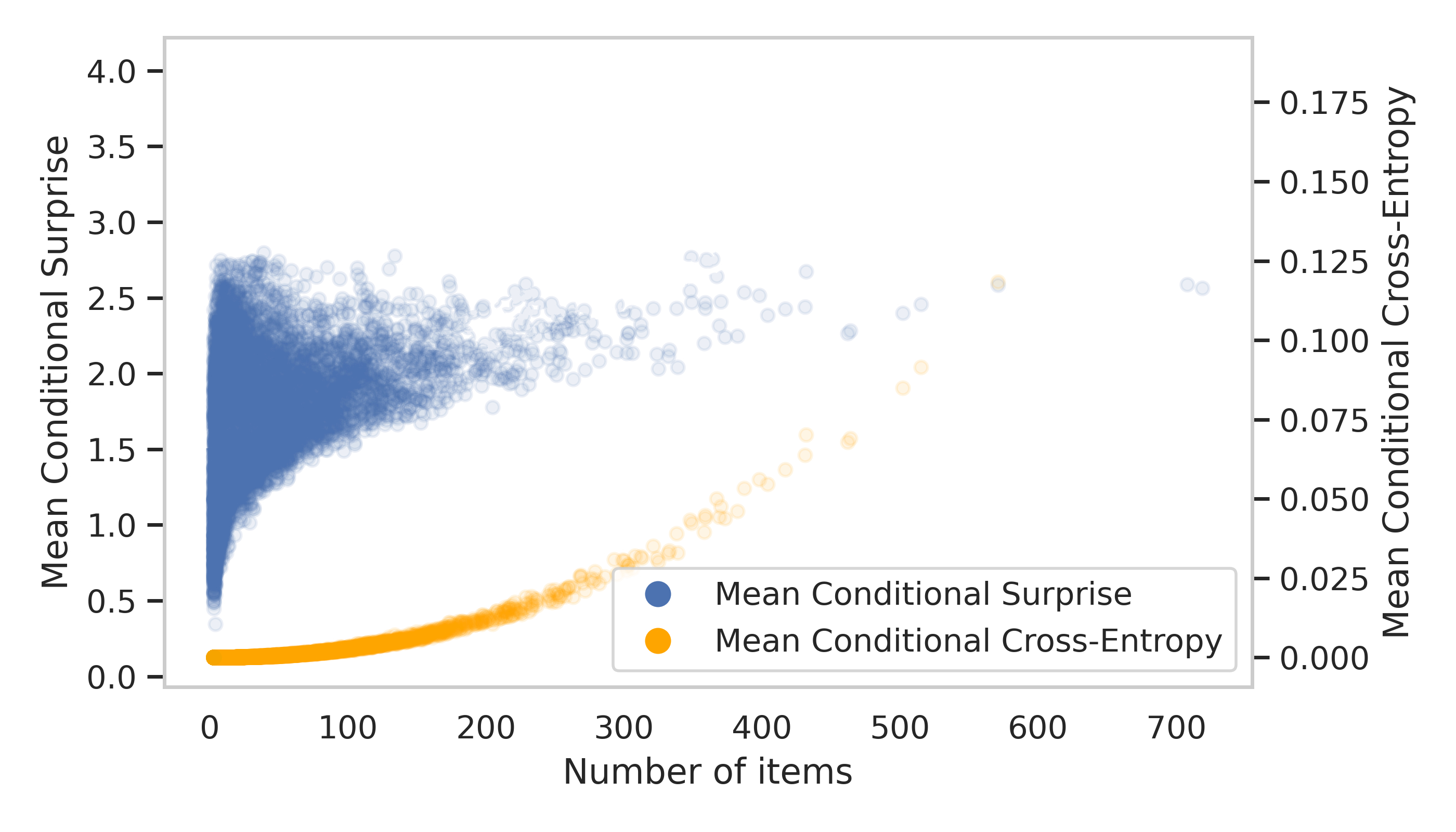}
  \vspace{-0.7cm}
  \caption{Comparison of $CS(u)$ and $\widetilde{CS}(u)$ against $|u|$ on the Netflix dataset}
   \label{fig:information_vs_ce}
  \vspace{-10pt} 
\end{figure}

We graphically inspect the relationship between candidate measures and $|u|$.
Figure \ref{fig:information_vs_ce} shows that $\widetilde{CS}(u)$ quadratically increases with the number of items, whereas $CS(u)$ tends to stabilize as $|u|$ increases.

A closer inspection reveals that $CS(u)$ still increases with $|u|$, but with a constant asymptotic behavior. This is consistent with the idea that users with fewer consumptions can either have very diverse consumptions or, on the contrary, exhibit narrow preferences. In contrast, users with more items, on average, tend to have more similar item coherence levels.

A potential concern is that the high variance of $CS(u)$ for users with small profiles (`cold-start' users) could be a statistical artifact rather than a true behavioral signal. We performed a Signal-to-Noise Ratio (SNR) analysis (see Appendix \ref{sec:snr_analysis}), which confirms that this is not the case. Even for the sparsest users, the true behavioral signal is much stronger than the measurement noise, confirming that our measure captures a genuine phenomenon even in challenging, data-scarce scenarios.

\paragraph{Correlation between measures.}

\begin{table}
  \centering
    \caption{Linear dependence between the information measures across datasets}
  
    \begin{tabular}{c||c|c}
    \toprule
        Dataset & Coefficient & $R^2$ \\
        \midrule
        ML 1M & 0.58 & 0.79 \\
        ML 10M & 0.55 & 0.86 \\
        Netflix Small & 0.49 & 0.67 \\
        Netflix & 0.6 & 0.84 \\
        Vis2Rec & -0.12 & 0.02 \\
        Tradesy & -0.66 & 0.28 \\
        Amazon Music & -0.66 & 0.43 \\
        Amazon Office & -0.68 & 0.44 \\
        Amazon Toys & -0.66 & 0.51\\
        \bottomrule
    \end{tabular}
    \label{tab:correlation}
  \vspace{-10pt}
\end{table}
Table \ref{tab:correlation} presents the linear dependence between $CS(u)$ and $S(u)$  across the tested datasets. 
As before, we see a clear distinction between the different domains. For movie datasets, we observe positive correlation coefficients ranging from 0.49 to 0.6, suggesting that in this domain, unsurprising users also happen to be coherent. The higher $R^2$ value for these datasets indicates that this relationship is quite significant and consistent.

In contrast, e-commerce datasets demonstrate strong negative correlations, with coefficients consistently between -0.66 and -0.68. This negative relationship implies that in online shopping contexts, the coherent users are also the most surprising ones. The moderate $R^2$ values suggest that while this inverse relationship is significant, it's not as deterministic as in the movie domain.

The Vis2Rec dataset stands out with a weak negative correlation (-0.12) and a very low $R^2$. This implies that for this dataset, there is only a very weak link between surprise and coherence.

These findings help us understand the data-specific challenges in each domain by revealing the unique relationships between user surprise and coherence. These domain-specific patterns underscore the importance of tailoring recommendation strategies to the unique characteristics of user behavior in each field, rather than applying a one-size-fits-all approach across different domains.

\subsection{Validating Coherence Measures}
For each dataset, we demonstrate the usefulness of our coherence measures by performing a logistic regression of the binary \texttt{Recall@20} metric denoted $Rec(u)$ on the oracle test information measures and the profile density, controlling for the used algorithm:
\begin{equation}
    \label{eq:oracle_regression}
\begin{aligned}
    Rec(u) \sim \sigma(algo+|u|\times S(i_{test})\times CS(i_{test}|u))
\end{aligned}
\end{equation}
We compute the average marginal effects (AME) of the variables and the McFadden's $R^2$ of the model, reporting the values in Table \ref{tab:oracle_regression_res}. The regressions have overall very high $R^2$ values, which shows the power of our regression\footnote{For logistic regression, McFadden $R^2>0.2$ is considered an excellent fit \cite{allison_measures_nodate}}. We additionally noticed that the marginal effects of the measures, in particular $CS(u)$, are as important as the algorithm's. The mostly negative coefficients highlight the sensitivity of the algorithms to the coherence of the test item with respect to the input items.

\begin{table}

        \caption{Marginal effects of the variables for equation \ref{eq:oracle_regression}. All values are significant at $p$-value $<.05$. For e.g. on ML 1M, the increase of 1 std. in $CS(i_{test}|u)$ causes the \texttt{Recall@20} to decrease of 22 points on average.}
    \label{tab:oracle_regression_res}
    \centering
        
    \begin{tabular}{c||c|c|c|c}
    \toprule
  Dataset & $|u|$ &$S(i_{test})$& $CS(i_{test}|u)$&$R^2$\\ \midrule
        ML 1M & \textcolor[RGB]{18,0,0}{-0.02} & \textcolor[RGB]{0,42,0}{0.02} & \textcolor[RGB]{217,0,0}{-0.22} & 0.34\\
        ML 10M & \textcolor[RGB]{18,0,0}{-0.02} & \textcolor[RGB]{0,128,0}{0.06} & \textcolor[RGB]{255,0,0}{-0.26} & 0.50\\
        Netflix Small & \textcolor[RGB]{0,42,0}{0.02} & \textcolor[RGB]{0,0,0}{-0.0} & \textcolor[RGB]{187,0,0}{-0.19} & 0.53\\
        Netflix & \textcolor[RGB]{48,0,0}{-0.05} & \textcolor[RGB]{39,0,0}{-0.04} & \textcolor[RGB]{0,42,0}{0.02} & 0.14\\
        Vis2Rec & \textcolor[RGB]{0,106,0}{0.05} & \textcolor[RGB]{18,0,0}{-0.02} & \textcolor[RGB]{167,0,0}{-0.17} & 0.4\\
        Tradesy & \textcolor[RGB]{0,64,0}{0.03} & \textcolor[RGB]{29,0,0}{-0.03} & \textcolor[RGB]{48,0,0}{-0.05} & 0.32\\
        Amazon Music & \textcolor[RGB]{0,85,0}{0.04} & \textcolor[RGB]{48,0,0}{-0.05} & \textcolor[RGB]{79,0,0}{-0.08} & 0.33\\
        Amazon Office & \textcolor[RGB]{0,42,0}{0.02} & \textcolor[RGB]{8,0,0}{-0.01} & \textcolor[RGB]{29,0,0}{-0.03} & 0.31\\
        Amazon Toys & \textcolor[RGB]{0,106,0}{0.05} & \textcolor[RGB]{29,0,0}{-0.03} & \textcolor[RGB]{89,0,0}{-0.09} & 0.35\\
        \bottomrule
    \end{tabular}
    \vspace{-0.7cm}
\end{table}
\begin{table*}[t]
	\centering
	\caption{Average marginal effect of the variables on the performance. Each value corresponds to the causal variation in \texttt{Recall@20} when the variable goes up by 1 unit of standard deviation.}
    \label{tab:AME}
    \begin{normalsize}   
\setlength{\tabcolsep}{3.5pt}  
\renewcommand{\arraystretch}{0.9}  
\begin{tabular}{c||ccccccccc||ccccccccc||ccccccccc}
\toprule
  & \multicolumn{9}{c||}{Conditional Surprise $\boldsymbol{CS(u})$} & \multicolumn{9}{c||}{Surprise $\boldsymbol{S(u)}$} & \multicolumn{9}{c}{Profile density $\boldsymbol{|u|}$} \\
\cmidrule(lr){2-28}
Algorithm & \rotatebox{90}{A. Music} & \rotatebox{90}{A. Office} & \rotatebox{90}{A. Toys} & \rotatebox{90}{ML 10M} & \rotatebox{90}{ML 1M} & \rotatebox{90}{Netflix} & \rotatebox{90}{Netflix S} & \rotatebox{90}{Tradesy} & \rotatebox{90}{Vis2Rec} & \rotatebox{90}{A. Music} & \rotatebox{90}{A. Office} & \rotatebox{90}{A. Toys} & \rotatebox{90}{ML 10M} & \rotatebox{90}{ML 1M} & \rotatebox{90}{Netflix} & \rotatebox{90}{Netflix S} & \rotatebox{90}{Tradesy} & \rotatebox{90}{Vis2Rec} & \rotatebox{90}{A. Music} & \rotatebox{90}{A. Office} & \rotatebox{90}{A. Toys} & \rotatebox{90}{ML 10M} & \rotatebox{90}{ML 1M} & \rotatebox{90}{Netflix} & \rotatebox{90}{Netflix S} & \rotatebox{90}{Tradesy} & \rotatebox{90}{Vis2Rec} \\
\midrule
UserKNN & \textcolor[RGB]{111,0,0}{-13} & \textcolor[RGB]{34,0,0}{-4} & \textcolor[RGB]{137,0,0}{-16} & \textcolor[RGB]{8,0,0}{-1} & \textcolor[RGB]{43,0,0}{-5} & \textcolor[RGB]{0,0,0}{0} & \textcolor[RGB]{16,0,0}{-2} & \textcolor[RGB]{103,0,0}{-12} & \textcolor[RGB]{111,0,0}{-13} & \textcolor[RGB]{59,0,0}{-7} & \textcolor[RGB]{24,0,0}{-3} & \textcolor[RGB]{85,0,0}{-10} & \textcolor[RGB]{0,4,0}{1} & \textcolor[RGB]{0,0,0}{0} & \textcolor[RGB]{0,0,0}{0} & \textcolor[RGB]{50,0,0}{-6} & \textcolor[RGB]{77,0,0}{-9} & \textcolor[RGB]{43,0,0}{-5} & \textcolor[RGB]{0,51,0}{12} & \textcolor[RGB]{0,8,0}{2} & \textcolor[RGB]{0,38,0}{9} & \textcolor[RGB]{0,0,0}{0} & \textcolor[RGB]{0,0,0}{0} & \textcolor[RGB]{0,0,0}{0} & \textcolor[RGB]{197,0,0}{-23} & \textcolor[RGB]{0,55,0}{13} & \textcolor[RGB]{0,29,0}{7} \\
ItemKNN & \textcolor[RGB]{163,0,0}{-19} & \textcolor[RGB]{93,0,0}{-11} & \textcolor[RGB]{197,0,0}{-23} & \textcolor[RGB]{0,0,0}{0} & \textcolor[RGB]{16,0,0}{-2} & \textcolor[RGB]{0,0,0}{0} & \textcolor[RGB]{0,46,0}{11} & \textcolor[RGB]{127,0,0}{-15} & \textcolor[RGB]{24,0,0}{-3} & \textcolor[RGB]{111,0,0}{-13} & \textcolor[RGB]{50,0,0}{-6} & \textcolor[RGB]{145,0,0}{-17} & \textcolor[RGB]{0,4,0}{1} & \textcolor[RGB]{0,0,0}{0} & \textcolor[RGB]{0,0,0}{0} & \textcolor[RGB]{50,0,0}{-6} & \textcolor[RGB]{85,0,0}{-10} & \textcolor[RGB]{0,0,0}{0} & \textcolor[RGB]{0,76,0}{18} & \textcolor[RGB]{0,25,0}{6} & \textcolor[RGB]{0,59,0}{14} & \textcolor[RGB]{0,0,0}{0} & \textcolor[RGB]{0,0,0}{0} & \textcolor[RGB]{0,0,0}{0} & \textcolor[RGB]{16,0,0}{-2} & \textcolor[RGB]{0,89,0}{21} & \textcolor[RGB]{0,25,0}{6} \\
WMF & \textcolor[RGB]{187,0,0}{-22} & \textcolor[RGB]{111,0,0}{-13} & \textcolor[RGB]{221,0,0}{-26} & \textcolor[RGB]{93,0,0}{-11} & \textcolor[RGB]{69,0,0}{-8} & \textcolor[RGB]{205,0,0}{-24} & \textcolor[RGB]{119,0,0}{-14} & \textcolor[RGB]{137,0,0}{-16} & \textcolor[RGB]{153,0,0}{-18} & \textcolor[RGB]{127,0,0}{-15} & \textcolor[RGB]{59,0,0}{-7} & \textcolor[RGB]{171,0,0}{-20} & \textcolor[RGB]{43,0,0}{-5} & \textcolor[RGB]{0,0,0}{0} & \textcolor[RGB]{0,17,0}{4} & \textcolor[RGB]{0,0,0}{0} & \textcolor[RGB]{103,0,0}{-12} & \textcolor[RGB]{24,0,0}{-3} & \textcolor[RGB]{0,85,0}{20} & \textcolor[RGB]{0,29,0}{7} & \textcolor[RGB]{0,64,0}{15} & \textcolor[RGB]{34,0,0}{-4} & \textcolor[RGB]{0,0,0}{0} & \textcolor[RGB]{16,0,0}{-2} & \textcolor[RGB]{0,51,0}{12} & \textcolor[RGB]{0,93,0}{22} & \textcolor[RGB]{0,64,0}{15} \\
LightGCN & \textcolor[RGB]{77,0,0}{-9} & \textcolor[RGB]{69,0,0}{-8} & \textcolor[RGB]{145,0,0}{-17} & \textcolor[RGB]{43,0,0}{-5} & \textcolor[RGB]{111,0,0}{-13} & \textcolor[RGB]{153,0,0}{-18} & \textcolor[RGB]{0,25,0}{6} & \textcolor[RGB]{85,0,0}{-10} & \textcolor[RGB]{137,0,0}{-16} & \textcolor[RGB]{50,0,0}{-6} & \textcolor[RGB]{43,0,0}{-5} & \textcolor[RGB]{111,0,0}{-13} & \textcolor[RGB]{93,0,0}{-11} & \textcolor[RGB]{0,21,0}{5} & \textcolor[RGB]{0,8,0}{2} & \textcolor[RGB]{137,0,0}{-16} & \textcolor[RGB]{69,0,0}{-8} & \textcolor[RGB]{34,0,0}{-4} & \textcolor[RGB]{0,34,0}{8} & \textcolor[RGB]{0,12,0}{3} & \textcolor[RGB]{0,42,0}{10} & \textcolor[RGB]{16,0,0}{-2} & \textcolor[RGB]{0,0,0}{0} & \textcolor[RGB]{0,0,0}{0} & \textcolor[RGB]{213,0,0}{-25} & \textcolor[RGB]{0,42,0}{10} & \textcolor[RGB]{0,38,0}{9} \\
RecVAE & \textcolor[RGB]{145,0,0}{-17} & \textcolor[RGB]{85,0,0}{-10} & \textcolor[RGB]{145,0,0}{-17} & \textcolor[RGB]{69,0,0}{-8} & \textcolor[RGB]{0,0,0}{0} & \textcolor[RGB]{205,0,0}{-24} & \textcolor[RGB]{111,0,0}{-13} & \textcolor[RGB]{93,0,0}{-11} & \textcolor[RGB]{137,0,0}{-16} & \textcolor[RGB]{85,0,0}{-10} & \textcolor[RGB]{50,0,0}{-6} & \textcolor[RGB]{93,0,0}{-11} & \textcolor[RGB]{59,0,0}{-7} & \textcolor[RGB]{59,0,0}{-7} & \textcolor[RGB]{0,21,0}{5} & \textcolor[RGB]{0,12,0}{3} & \textcolor[RGB]{69,0,0}{-8} & \textcolor[RGB]{24,0,0}{-3} & \textcolor[RGB]{0,64,0}{15} & \textcolor[RGB]{0,21,0}{5} & \textcolor[RGB]{0,42,0}{10} & \textcolor[RGB]{43,0,0}{-5} & \textcolor[RGB]{0,0,0}{0} & \textcolor[RGB]{24,0,0}{-3} & \textcolor[RGB]{0,0,0}{0} & \textcolor[RGB]{0,64,0}{15} & \textcolor[RGB]{0,68,0}{16} \\
EASE & \textcolor[RGB]{197,0,0}{-23} & \textcolor[RGB]{111,0,0}{-13} & \textcolor[RGB]{221,0,0}{-26} & \textcolor[RGB]{34,0,0}{-4} & \textcolor[RGB]{137,0,0}{-16} & \textcolor[RGB]{213,0,0}{-25} & \textcolor[RGB]{103,0,0}{-12} & \textcolor[RGB]{187,0,0}{-22} & \textcolor[RGB]{153,0,0}{-18} & \textcolor[RGB]{137,0,0}{-16} & \textcolor[RGB]{69,0,0}{-8} & \textcolor[RGB]{171,0,0}{-20} & \textcolor[RGB]{111,0,0}{-13} & \textcolor[RGB]{0,25,0}{6} & \textcolor[RGB]{0,12,0}{3} & \textcolor[RGB]{0,0,0}{0} & \textcolor[RGB]{119,0,0}{-14} & \textcolor[RGB]{24,0,0}{-3} & \textcolor[RGB]{0,89,0}{21} & \textcolor[RGB]{0,29,0}{7} & \textcolor[RGB]{0,68,0}{16} & \textcolor[RGB]{34,0,0}{-4} & \textcolor[RGB]{0,0,0}{0} & \textcolor[RGB]{0,0,0}{0} & \textcolor[RGB]{0,0,0}{0} & \textcolor[RGB]{0,128,0}{30} & \textcolor[RGB]{0,59,0}{14}\\
\bottomrule
\end{tabular}
\end{normalsize}
\vspace{-10pt}
\end{table*}
\subsection{Impact on Performance}

We now directly estimate how our measures computed on the train set impact RS performance. Figure \ref{fig:side-by-side} shows the relation of the \texttt{Recall@20} to $CS(u)$. This graph reveals several important insights:
\begin{itemize}[noitemsep, leftmargin=0.3cm]
    \item There is a clear negative correlation between $CS(u)$ and recommendation performance for all algorithms.
    \item The performance gap between different algorithms is most pronounced for coherent users, i.e. for users with low $CS(u)$ values.
    \item As $CS(u)$ increases, the performance of all algorithms converges to a similarly low level.
\end{itemize}
Notably, the convergence of algorithm performance for high $CS(u)$ values suggests that for highly incoherent users, the choice of algorithm becomes less important. This observation has significant implications for the design and deployment of recommender systems. It indicates that \textbf{most gains in overall performance primarily come from improvements in recommendations for coherent users}. For incoherent users, even sophisticated algorithms struggle to outperform simpler approaches.

To get the true marginal effect independent of other variables, we estimate the model:
\begin{equation}
    \label{eq:main_regression}
    Rec(u) \sim \sigma(|u|\times S(u) \times CS(u))
\end{equation}
We perform one regression on each dataset and algorithm pair. We model the variability in $\log(p_i)$ and $\log(p_{i|j})$ for a given user by using SIMEX with a variance estimated on each user set of interactions. 

Empirically, this yields to a model with much more statistically significant effects, and with a larger effect norm.

The AME for the regression of each dataset and algorithm are reported in Table \ref{tab:AME}, showing that the most important effects come from profile density and $CS(u)$. 
The AME of $|u|$ is non-significant or negative for movie sets while being positive for e-commerce. This can be explained by the difference in consumption density between movie and e-commerce datasets. Since e-commerce profiles are more sparse, each new item adds useful information for RS. On the other hand, adding items to already dense profiles only adds complexity.

\textbf{The $S(u)$ measure is the less impactful variable}, meaning RS adapt (to some degree) to users with niche tastes. Mean Surprise still holds a negative effect on e-commerce sets. This could be explained by the fact that these datasets have a higher mean $S(u)$. When a user deviates from the popular items, they would buy very rare items, which are not well-modeled by the algorithm.

\textbf{Mean Conditional Surprise $CS(u)$ greatly impacts performance negatively in most scenarios}. This highlights the importance of the measure in quantifying the difficulty of a user. The distinction between e-commerce and movie datasets is not as clear as for the previous graphs, showing the cross-domain applicability of the measure.

\subsection{Applications in Algorithmic Analysis}
\paragraph{Stratified evaluation}
A key application of our measures is to enable a more granular and insightful form of algorithm assessment. Instead of relying on a single, aggregate performance metric, practitioners can report scores across user segments defined by their coherence. Figure \ref{fig:side-by-side} is a clear example of this methodology, revealing an insight that an overall average would hide: all tested algorithms universally fail for incoherent users. This stratified view not only deepens our understanding but also motivates treating these user segments differently in a production environment --- for instance, by routing them to bespoke models to optimize system-wide performance and efficiency. The following subsections explore such applications in detail.

This approach moves the evaluation beyond simply asking \textit{"which model is better?"} to asking \textit{"which model is better, for whom, and why?"}. It allows for the creation of more challenging benchmarks focused on specific user types (e.g., incoherent users) and provides a deeper understanding of model strengths and weaknesses. This methodology is the foundation for the more advanced analytical and system design applications discussed below.

\paragraph{Coherence Reproduction as an Analytical Tool.}
A good recommendation system should not only predict relevant items but also generate a set of suggestions that align with a user's intrinsic consumption patterns. While traditional metrics like \texttt{Recall@K} measure item-level correctness, they do not capture this "behavioral alignment." We propose evaluating this alignment by measuring how well a model's predictions reproduce the coherence and surprise levels of a user's input profile.

\begin{figure}[t]
    \centering
     \includegraphics[width=\linewidth]{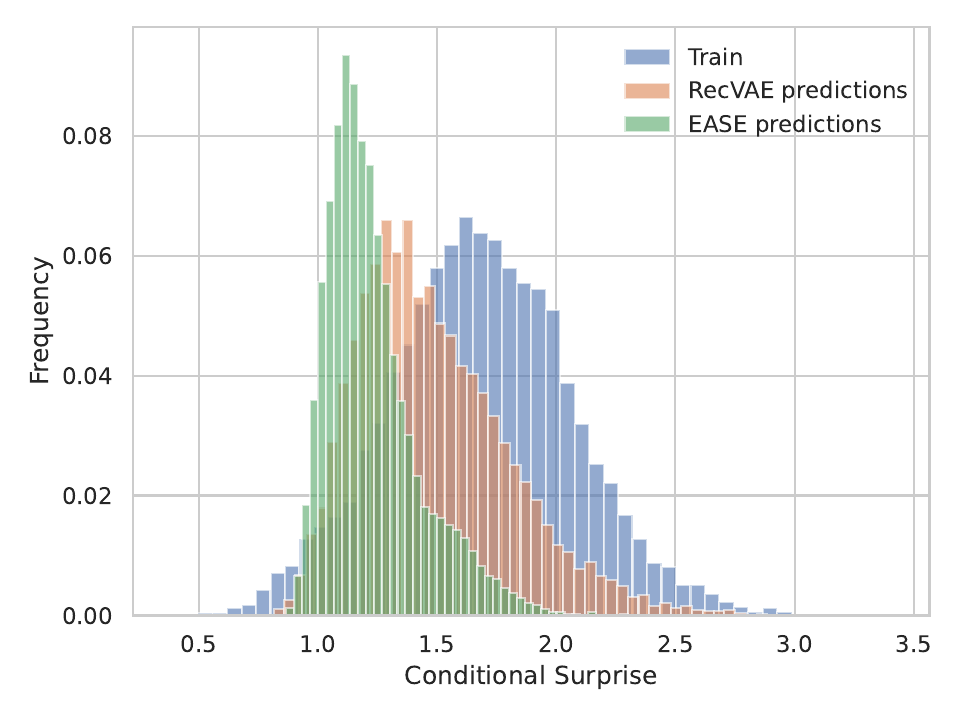}
     \vspace{-0.7cm}
    \caption{ Distributions of $CS(u)$ for ML 10M (Train) set and predictions of both RecVAE and EASE}
\label{fig:coherence_dist_repro}
    \vspace{-0.75cm}
\end{figure}

At an aggregate level, we can visualize this alignment by comparing distributions. Figure \ref{fig:coherence_dist_repro} provides an illustrative example, comparing the distribution of $CS(u)$ from the ML-10M training set against the distributions from the recommendation sets of \texttt{EASE} and \texttt{RecVAE}. We observe that \texttt{EASE}, despite its higher recall, tends to recommend item sets that are, on average, more coherent (lower $CS$) than the user's actual history. In contrast, \texttt{RecVAE}'s output distribution more closely matches the input, suggesting it is better at capturing the diversity of user coherence patterns, despite lower recall.

\begin{table*}
\caption{Correlation between $S(u)$ and $S(\hat{u})$; and $CS(u)$ and $CS(\hat{u})$, where $\hat{u}$ is the predicted item set for $u$. The higher the correlation, the better the RS can reproduce the coherence level of its inputs.}
\vspace{-0.3cm}
\label{tab:correlation_user}
\begin{normalsize}   
\setlength{\tabcolsep}{3.5pt}  
\renewcommand{\arraystretch}{0.9}  
\begin{tabular}{l||ccccccccc|c||ccccccccc|c}
\toprule
 & \multicolumn{10}{c||}{Surprise correlation} & \multicolumn{10}{c}{Conditional Surprise correlation} \\
\cmidrule(lr){2-21}
 & \rotatebox{90}{ML 1M} & \rotatebox{90}{ML 10M} & \rotatebox{90}{Netflix S} & \rotatebox{90}{Netflix} & \rotatebox{90}{Vis2rec} & \rotatebox{90}{Tradesy} & \rotatebox{90}{A. Music} & \rotatebox{90}{A. Office} & \rotatebox{90}{A. Toys} & \rotatebox{90}{\textbf{Mean}} & \rotatebox{90}{ML 1M} & \rotatebox{90}{ML 10M} & \rotatebox{90}{Netflix S} & \rotatebox{90}{Netflix} & \rotatebox{90}{Vis2rec} & \rotatebox{90}{Tradesy} & \rotatebox{90}{A. Music} & \rotatebox{90}{A. Office} & \rotatebox{90}{A. Toys} & \rotatebox{90}{\textbf{Mean}} \\
\midrule
MostPop & \textcolor[RGB]{7,0,0}{-.02} & \textcolor[RGB]{0,0,0}{.0} & \textcolor[RGB]{13,0,0}{-.04} & \textcolor[RGB]{16,0,0}{-.05} & \textcolor[RGB]{125,0,0}{-.38} & \textcolor[RGB]{79,0,0}{-.24} & \textcolor[RGB]{111,0,0}{-.34} & \textcolor[RGB]{93,0,0}{-.28} & \textcolor[RGB]{135,0,0}{-.41} & \textcolor[RGB]{65,0,0}{-.2} & \textcolor[RGB]{0,24,0}{.15} & \textcolor[RGB]{0,29,0}{.18} & \textcolor[RGB]{0,24,0}{.15} & \textcolor[RGB]{0,8,0}{.05} & \textcolor[RGB]{0,49,0}{.3} & \textcolor[RGB]{2,0,0}{-.01} & \textcolor[RGB]{145,0,0}{-.44} & \textcolor[RGB]{37,0,0}{-.11} & \textcolor[RGB]{0,3,0}{.02} & \textcolor[RGB]{0,4,0}{.03} \\
UserKNN & \textcolor[RGB]{69,0,0}{-.21} & \textcolor[RGB]{23,0,0}{-.07} & \textcolor[RGB]{27,0,0}{-.08} & \textcolor[RGB]{0,0,0}{.0} & \textcolor[RGB]{0,72,0}{.44} & \textcolor[RGB]{0,64,0}{.39} & \textcolor[RGB]{0,93,0}{.57} & \textcolor[RGB]{0,72,0}{.44} & \textcolor[RGB]{0,72,0}{.44} & \textcolor[RGB]{0,34,0}{.21} & \textcolor[RGB]{0,3,0}{.02} & \textcolor[RGB]{0,6,0}{.04} & \textcolor[RGB]{0,24,0}{.15} & \textcolor[RGB]{0,0,0}{.0} & \textcolor[RGB]{18,0,0}{-.06} & \textcolor[RGB]{0,8,0}{.05} & \textcolor[RGB]{0,106,0}{\textbf{.65}} & \textcolor[RGB]{0,93,0}{\textbf{.57}} & \textcolor[RGB]{0,87,0}{\textbf{.53}} & \textcolor[RGB]{0,36,0}{.22} \\
ItemKNN & \textcolor[RGB]{105,0,0}{-.32} & \textcolor[RGB]{165,0,0}{-.5} & \textcolor[RGB]{227,0,0}{-.69} & \textcolor[RGB]{161,0,0}{-.49} & \textcolor[RGB]{39,0,0}{-.12} & \textcolor[RGB]{0,54,0}{.33} & \textcolor[RGB]{0,72,0}{.44} & \textcolor[RGB]{0,59,0}{.36} & \textcolor[RGB]{0,67,0}{.41} & \textcolor[RGB]{23,0,0}{-.07} & \textcolor[RGB]{98,0,0}{-.3} & \textcolor[RGB]{129,0,0}{-.39} & \textcolor[RGB]{75,0,0}{-.23} & \textcolor[RGB]{0,4,0}{.03} & \textcolor[RGB]{0,44,0}{.27} & \textcolor[RGB]{0,29,0}{\textbf{.18}} & \textcolor[RGB]{0,90,0}{.55} & \textcolor[RGB]{0,93,0}{\textbf{.57}} & \textcolor[RGB]{0,83,0}{.51} & \textcolor[RGB]{0,21,0}{.13} \\
WMF & \textcolor[RGB]{0,88,0}{.54} & \textcolor[RGB]{0,85,0}{.52} & \textcolor[RGB]{0,106,0}{.65} & \textcolor[RGB]{0,113,0}{.69} & \textcolor[RGB]{0,123,0}{\textbf{.75}} & \textcolor[RGB]{0,70,0}{\textbf{.43}} & \textcolor[RGB]{0,100,0}{\textbf{.61}} & \textcolor[RGB]{0,92,0}{\textbf{.56}} & \textcolor[RGB]{0,80,0}{.49} & \textcolor[RGB]{0,95,0}{.58} & \textcolor[RGB]{0,42,0}{.26} & \textcolor[RGB]{0,37,0}{.23} & \textcolor[RGB]{0,31,0}{.19} & \textcolor[RGB]{0,75,0}{.46} & \textcolor[RGB]{0,75,0}{.46} & \textcolor[RGB]{95,0,0}{-.29} & \textcolor[RGB]{0,41,0}{.25} & \textcolor[RGB]{0,37,0}{.23} & \textcolor[RGB]{0,60,0}{.37} & \textcolor[RGB]{0,39,0}{.24} \\
LightGCN & \textcolor[RGB]{0,121,0}{\textbf{.74}} & \textcolor[RGB]{0,111,0}{.68} & \textcolor[RGB]{0,70,0}{.43} & \textcolor[RGB]{0,128,0}{\textbf{.78}} & \textcolor[RGB]{0,106,0}{.65} & \textcolor[RGB]{0,39,0}{.24} & \textcolor[RGB]{0,65,0}{.4} & \textcolor[RGB]{0,90,0}{.55} & \textcolor[RGB]{0,88,0}{\textbf{.54}} & \textcolor[RGB]{0,92,0}{.56} & \textcolor[RGB]{0,101,0}{\textbf{.62}} & \textcolor[RGB]{0,77,0}{.47} & \textcolor[RGB]{0,3,0}{.02} & \textcolor[RGB]{0,103,0}{.63} & \textcolor[RGB]{0,95,0}{\textbf{.58}} & \textcolor[RGB]{0,23,0}{.14} & \textcolor[RGB]{0,75,0}{.46} & \textcolor[RGB]{0,87,0}{.53} & \textcolor[RGB]{0,77,0}{.47} & \textcolor[RGB]{0,70,0}{\textbf{.43}} \\
RecVAE & \textcolor[RGB]{0,90,0}{.55} & \textcolor[RGB]{0,123,0}{\textbf{.75}} & \textcolor[RGB]{0,110,0}{\textbf{.67}} & \textcolor[RGB]{0,126,0}{.77} & \textcolor[RGB]{0,101,0}{.62} & \textcolor[RGB]{0,82,0}{.5} & \textcolor[RGB]{0,75,0}{.46} & \textcolor[RGB]{0,85,0}{.52} & \textcolor[RGB]{0,88,0}{\textbf{.54}} & \textcolor[RGB]{0,98,0}{\textbf{.60}} & \textcolor[RGB]{0,93,0}{.57} & \textcolor[RGB]{0,101,0}{\textbf{.62}} & \textcolor[RGB]{0,87,0}{\textbf{.53}} & \textcolor[RGB]{0,118,0}{\textbf{.72}} & \textcolor[RGB]{0,44,0}{.27} & \textcolor[RGB]{0,11,0}{.07} & \textcolor[RGB]{0,44,0}{.27} & \textcolor[RGB]{0,55,0}{.34} & \textcolor[RGB]{0,65,0}{.4} & \textcolor[RGB]{0,69,0}{.42} \\
EASE & \textcolor[RGB]{0,92,0}{.56} & \textcolor[RGB]{0,75,0}{.46} & \textcolor[RGB]{0,100,0}{.61} & \textcolor[RGB]{0,113,0}{.69} & \textcolor[RGB]{0,98,0}{.6} & \textcolor[RGB]{0,70,0}{\textbf{.43}} & \textcolor[RGB]{0,77,0}{.47} & \textcolor[RGB]{0,72,0}{.44} & \textcolor[RGB]{0,77,0}{.47} & \textcolor[RGB]{0,87,0}{.53} & \textcolor[RGB]{0,65,0}{.4} & \textcolor[RGB]{0,37,0}{.23} & \textcolor[RGB]{0,46,0}{.28} & \textcolor[RGB]{0,88,0}{.54} & \textcolor[RGB]{0,88,0}{.54} & \textcolor[RGB]{0,9,0}{.06} & \textcolor[RGB]{0,83,0}{.51} & \textcolor[RGB]{0,67,0}{.41} & \textcolor[RGB]{0,78,0}{.48} & \textcolor[RGB]{0,62,0}{.38} \\
\bottomrule
\end{tabular}
\vspace{-0.4cm}
\end{normalsize}
\end{table*}
\begin{table}[!h]
    \vspace{0.1cm}
    \caption{\texttt{Recall@20} on the coherent users of Netflix, for the Vanilla models and the specialized ones, trained on a small coherent subset.}
    \label{tab:my_label}
    \centering
    \vspace{-0.3cm}
    \begin{tabular}{c|ccccc}
    \toprule
        & ItemKNN & LGCN & WMF & RVAE & EASE \\ \midrule
      Vanilla & 0.0 & 33.8 & 47.4 & 46.0 & 53.0 \\
    \textbf{Spec.} & \textbf{3.2} & \textbf{39.3} & \textbf{49.2} & \textbf{47.8} & \textbf{56.0}\\
    \bottomrule
    \end{tabular}
    \vspace{-0.5cm}
\end{table}

To quantify this phenomenon at a user level, we compute the Pearson correlation between the coherence measures of a user's input profile $u$ and their recommended set $\hat{u}$. Table \ref{tab:correlation_user} presents these correlations across all datasets and algorithms. The results reveal that most algorithms are reasonably effective at reproducing a user's general preference for popular versus niche items (strong positive correlation for $S(u)$). However, they are far less successful at preserving the internal consistency of a user's choices (much weaker correlation for $CS(u)$). While models can capture first-order popularity, they struggle to replicate the more complex, second-order relationships. Deep learning methods like \texttt{LightGCN} and \texttt{RecVAE} show a slightly stronger correlation for $CS(u)$, suggesting they may better model these complex interactions.

These findings suggest that \textbf{Coherence Preservation}, measured as the user-wise correlation between input and output $CS(u)$, should be considered a complementary metric for RS analysis. It reveals algorithmic strengths that accuracy metrics miss and provides a valuable tool for practitioners. For example, it can help identify models that are better suited for tasks where preserving user diversity and avoiding taste homogenization are key objectives, such as in discovery-oriented or long-tail recommendation scenarios.

\paragraph{User Segmentation for Specialized Models.}

Beyond analysis, our measures provide a practical tool for system design. We highlight their utility by investigating whether segmenting users by their coherence can enable specialized models to improve performance on targeted user groups. We conduct a proof-of-concept experiment on the large-scale Netflix dataset, which is sufficiently large to allow for meaningful segmentation. We first identify a train and a test set composed of the most coherent users (those in the lowest decile of the $CS(u)$ distribution). We then train and test our algorithms on these new "coherent" subsets.

Table \ref{tab:my_label} compares the \texttt{Recall@20} of these specialized models against the vanilla models (trained on all data) when evaluated on this specific coherent segment. Despite using significantly less training data, the specialized models consistently achieve better performance across all tested algorithms.

The results in Table \ref{tab:my_label} provide a valuable insight into modeling coherent users. The consistent, albeit modest, performance improvement suggests that for this predictable user segment, the noise introduced by incoherent user data in a large training set can be more detrimental than the benefit of sheer data volume. This validates $CS(u)$ as an effective tool for identifying high-signal user segments where data quality—in the sense of behavioral consistency—is a critical factor for achieving optimal performance.

\section{Practical Implications and Future Work}
Our study focuses on foundational CF models with implicit feedback, which opens several avenues for future work. First, while we provide the tools to identify incoherent users, designing novel architectures that effectively serve this challenging segment remains an open problem. Second, extending this framework to other data modalities, such as explicit ratings or session-based data, is a valuable next step. Finally, applying our coherence analysis to the new generation of LLM-based recommenders could yield important insights into their behavioral patterns and failure modes.

The $S(u)$ and $CS(u)$ measures are not just analytical tools for future and past works; they can be integrated into production systems to inform strategy and improve user experience. We suggest the following high-level applications:



 \noindent \textbf{Robust Evaluation and Benchmarking.} Go beyond aggregate metrics by adopting stratified evaluation. Reporting performance on distinct segments of coherent and incoherent users should become standard practice for A/B testing and model selection, as it uncovers critical failure modes that overall averages can hide.
 
 \noindent \textbf{Adaptive Personalization Strategies.} Use user coherence as a signal to dynamically switch recommendation modes. For highly coherent users (low $CS(u)$), an "exploit" strategy focused on deep personalization may be optimal. Conversely, for incoherent users (high $CS(u)$), where prediction is likely to fail, the system could switch to a robust "explore" strategy, recommending diverse and popular items to help the user build a more established taste profile.

 \noindent \textbf{Informing the Cold-Start Experience.} Use a user's initial coherence score after their first few interactions as a feature. A new user exhibiting highly incoherent behavior might be a candidate for a different onboarding experience, such as an explicit preference elicitation step, to help the system learn their interests more quickly and effectively.
 \vspace{-0.25cm}
\section{Conclusion}
We introduced two information-theoretic measures, Mean Surprise ($S(u)$) and Mean Conditional Surprise ($CS(u)$), to provide a new framework for analyzing recommender systems. Our extensive experiments show that these measures effectively quantify distinct aspects of user behavior: $S(u)$ characterizes a user's preference for popular versus niche items, while the domain-agnostic $CS(u)$ measures a user's internal profile coherence.

Our central finding is that recommendation performance is strongly tied to user coherence, with performance gains from complex models largely concentrated on "coherent" users. As demonstrated, this insight enables several practical applications. Our measures can be used for more robust, \textbf{stratified evaluation}; to analyze\textbf{ behavioral alignment} by measuring "Coherence Preservation"; and to inform system design through \textbf{user segmentation}, for which we provided a successful proof-of-concept with specialized models. This work highlights the importance of user coherence modeling, providing practical tools to develop more efficient, adaptive, and comprehensible recommender systems.\\
\textbf{Acknowledegments: }This work was partly supported by the SHARP ANR project ANR-23-PEIA-0008 in the context of the France 2030 program.It was made possible by the use of the FactoryIA supercomputer, financially supported by the Ile-DeFrance Regional Council.

\bibliographystyle{plainnat}

{
\small
\bibliography{ref}
}

\appendix
\section{Properties of the Measures}
\subsection{Orthogonality of the measures}
Since $\log(p_{i|i}) = 0$, effectively, the pairs $(i,i)$ do not intervene in the definition of $SC(u)$. In fact, if we denote by $PS(u)$ the \textbf{Mean Pair Surprise} by replacing $p_{i|j}$ by $p_{i,j}$ in the definition of $CS(u)$, then we have:
\begin{equation}
    CS(u) = PS(u) - S(u)
\end{equation}
Effectively, we remove the effect of the surprise at the first order from the surprise of the pairs.

\subsection{Proof of Proposition 1}
\label{sup:proof1}
Let $p_{ui} = \mathbb{P}(x_{ui}=1||u|>0)$. First, consider that:
\begin{align*}
      \mathbb{E}_{\pi_u^{\geq 1}}[\widetilde{S}(u)] &= \frac{1}{m}\sum_{i=1}^m \log(p_i^*)\mathbb{E}_{\pi_u^{\geq 1}}[x_{ui}]\\
      &=\frac{1}{m}\sum_{i=1}^m \log(p_i^*)p_{ui}
\end{align*}
since $x_{ui}$ is a Bernoulli variable.
Then for any user $u$ with $|u| > 0$:
\begin{align*}
    \mathbb{E}_{\pi_u^{\geq 1}}[S(u)] &= -\sum_i^m \log(p_i^*) \mathbb{E}_{\pi_u^{\geq 1}}\left[\frac{x_{ui}}{|u|}\right]\\
    &= -\sum_i^m \log(p_i^*) \mathbb{E}_{x_{ui}}\left[\mathbb{E}_{\pi_u^{\geq 1}}\left[\frac{x_{ui}}{|u|}\bigg|x_{ui}\right]\right]\\
    \end{align*}
    \begin{align*}
    &=-\sum_i^m \log(p_i^*) p_{ui}\mathbb{E}_{\pi_u^{\geq 1}}\left[\frac{x_{ui}}{|u|}\bigg|x_{ui}=1\right]\\
    &=-\sum_i^m \log(p_i^*)p_{ui}\mathbb{E}_{\pi_u^{\geq 1}}\left[\frac{1}{|u|}\bigg|x_{ui}=1\right]
\end{align*}
where the second line is a consequence of the law of iterated expectations. Then, since $x\mapsto 1/x$ is convex on $\mathbb{R}^*_{+}$, by Jensen inequality:
\begin{align*}
      \mathbb{E}_{\pi_u^{\geq 1}}[S(u)] &\geq -\sum_i^m \log(p_i^*) p_{ui}\frac{1}{\mathbb{E}_{\pi_u^{\geq 1}}[|u|]}\\
      &\geq \mathbb{E}_{\pi_u^{\geq 1}}[\widetilde{S}(u)]\frac{m}{\mathbb{E}_{\pi_u^{\geq 1}}[|u|]}
\end{align*}
which directly gives the left-hand side of Proposition 1. The right-hand-side follows from the fact that $|u| = \sum_i x_{ui}$. So, for any random events $\omega_0, \omega_1$ that only differ in $x_{ui}(\omega_0)=0$ and $x_{ui}(\omega_1)=1$, we have $\frac{1}{|u|}(\omega_0)\geq\frac{1}{|u|}(\omega_1)$. Taking the expected value from both sides gives:
\begin{align*}
     \mathbb{E}_{\pi_u^{\geq 1}}[S(u)] &\leq -\sum_i^m \log(p_i^*)p_{ui}\mathbb{E}_{\pi_u^{\geq 1}}\left[\frac{1}{|u|}\right]\\
     &\leq -\mathbb{E}_{\pi_u^{\geq 1}}\left[\frac{1}{|u|}\right]\sum_i^m \log(p_i^*)p_{ui} \\
     &\leq \mathbb{E}_{\pi_u^{\geq 1}}\left[\frac{m}{|u|}\right]\mathbb{E}_{\pi_u^{\geq 1}}[\widetilde{S}(u)]
     \quad 
     \square
\end{align*}
For $CS(u)$, the convexity of $x\mapsto1/x^2$ on $\mathbb{R}^*_+$ and the fact that $\mathbb{E}_{\pi_u^{\geq 1}}\left[\frac{1}{|u|^2}|x_{ui}=1\right]\leq \mathbb{E}_{\pi_u^{\geq 1}}\left[\frac{1}{|u|^2}\right]$ proves the bounds.
\begin{figure}[t]
    \centering
    \includegraphics[width=0.8\linewidth]{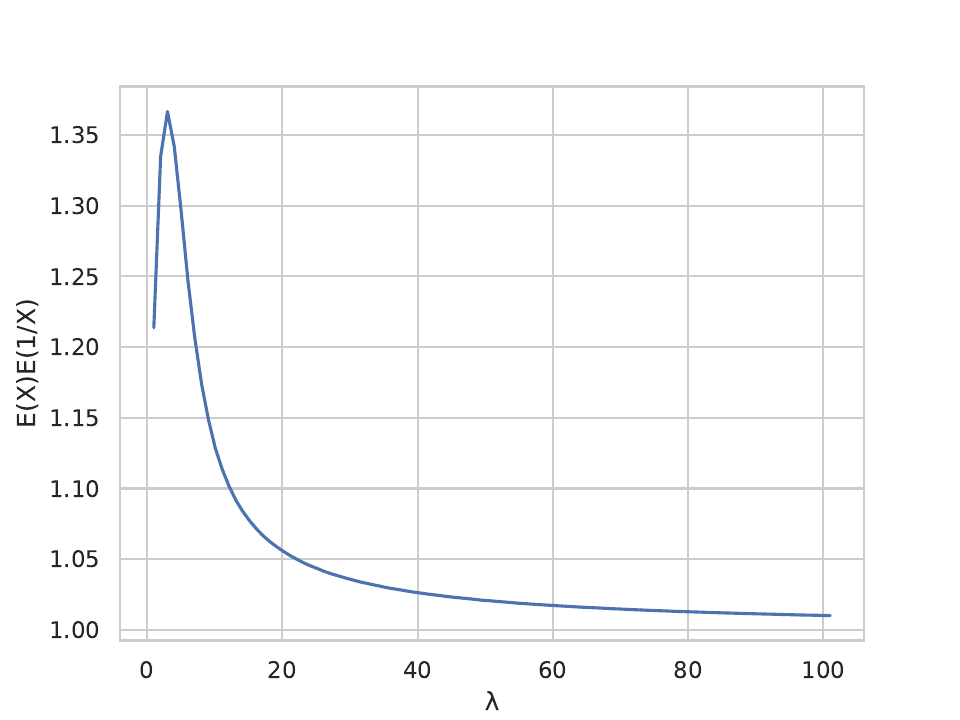}
    \caption{Monte-Carlo estimation of  $\mathbb{E}_{X\geq1}[X]\mathbb{E}_{X\geq1}\left[\frac{1}{X}\right]$}
    \label{fig:estimation1}
\end{figure}
\subsection{Discussion about the Poisson model for $|u|$}
\label{sup:PoissonModel}
The Poison distribution, also known as "the law of rare events", is an adapted model to count the frequency of events that occur rarely. In particular, in recommendation data, users consume only a small fraction of the possible items, giving a motivation for Poisson modelization. Moreover, if we choose a finer description of the user's choices, for example, assigning known oracle probability $p_{ui}$ of observing the item $i$ in the user's $u$ set, then $\pi_u$ becomes, by definition, a multivariate Bernoulli distribution. Therefore, $|u| = \sum x_{ui}$ is, by definition, a Poisson-Binomial variable, which is well-approximated by a Poisson distribution of parameter $\lambda=\sum p_{ui}$, in virtue of Le Cam's theorem \cite{lecam}.

\begin{figure}[t]
    \centering
    \includegraphics[width=0.8\linewidth]{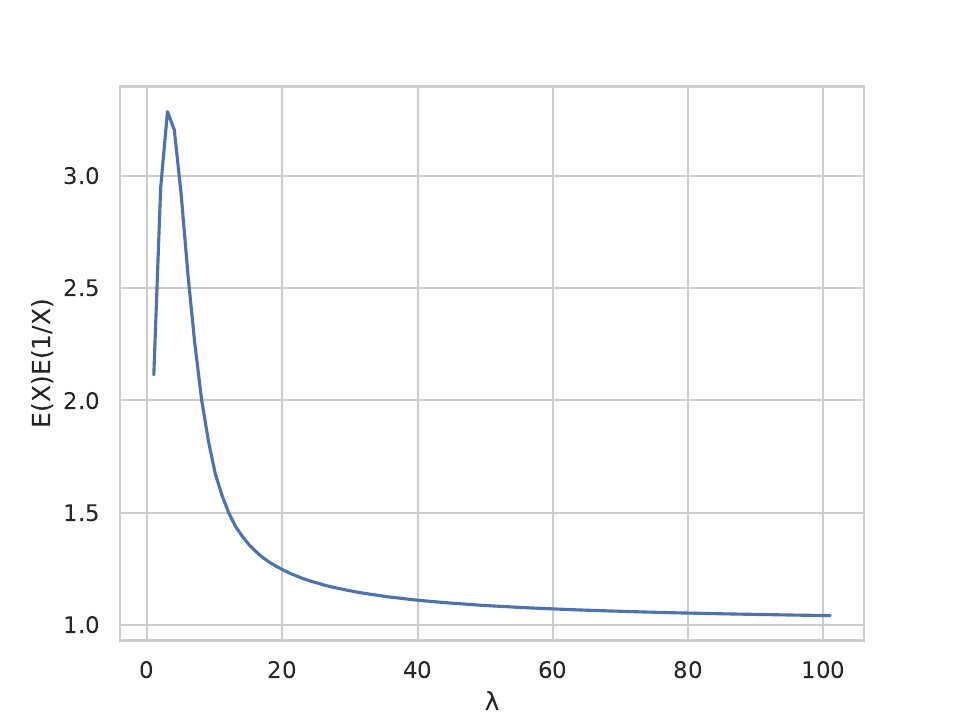}
    \caption{Monte Carlo estimation of $\mathbb{E}_{X\geq1}[X^2]\mathbb{E}_{X\geq1}\left[\frac{1}{X^2}\right]$}
    \label{fig:estimation2}
\end{figure}
\subsection{Proof of Proposition 2}
\label{sup:prop2}
If $X$ is a poison variable of parameter $\lambda >0$, then : 
\begin{equation*}
    \forall k \in \mathbb{N}, \mathbb{P}[X=k] = \frac{e^{-\lambda} \lambda^k}{k!}
\end{equation*}
which yields:
\begin{align*}
     \forall k \in \mathbb{N}, \mathbb{P}[X=k|X>0] &= \frac{\mathbb{P}[X>0|X=k] \mathbb{P}[X=k]}{1-\mathbb{P}[X=0]}\\
     &= \frac{\mathds{1}(k>0)}{1-e^{-\lambda}}\mathbb{P}[X=k]\\
     &= \mathds{1}(k>0) \frac{e^{-\lambda}\lambda^k}{(1-e^{-\lambda})k!}
\end{align*}
Then, we can consider the fact that  we have:
\begin{equation*}
    \forall k \in \mathbb{N}^*, \frac{1}{k} \leq \frac{2}{k+1}
\end{equation*}
Taking the expectancy conditioned on $X>0$ (i.e. $X \geq 1$ ) gives us:
\begin{align*}
    \mathbb{E}_{X\geq1}\left[\frac{1}{X}\right]&\leq   \mathbb{E}_{X\geq1}\left[\frac{2}{X+1}\right]\\
    &\leq 2\sum_{k=1}^\infty \frac{\mathbb{P}[X=k|X>0]}{k+1} \\
    &\leq 2\sum_{k=1}^\infty\frac{e^{-\lambda}}{1-e^{-\lambda}}\frac{\lambda^k}{(k+1)k!}\\
    &\leq \frac{2e^{-\lambda}}{1-e^{-\lambda}}\frac{1}{\lambda}\sum_{k=1}^\infty \frac{\lambda^{k+1}}{(k+1)!}\\
    &\leq \frac{2e^{-\lambda}}{1-e^{-\lambda}}\frac{1}{\lambda}(e^{\lambda} -1-\lambda)\\
    &\leq \frac{2(1-e^{-\lambda})}{\lambda}\\
    &\leq \frac{2}{\mathbb{E}_{\geq1}[X]} \quad \square
\end{align*}
To get a similar bound for $CS(u)$, i.e bounding $\mathbb{E}_{X\geq1}[X^2]\mathbb{E}_{X\geq1}\left[\frac{1}{X^2}\right]$ we first find a constant $K$ such that :
\begin{align*}
    \forall k \in \mathbb{N}^*, \frac{1}{k^2} &\leq \frac{K}{(k+1)(k+2)}\\
    0 &\leq(1-\frac{1}{K})k^2 - \frac{3}{K}k -\frac{2}{K}
\end{align*}
The biggest root of the RHS is given by $\frac{3+\sqrt{1+8K}}{2K-2}$, which is smaller than 1 for $K\geq 6$. The rest of the proof follows the same calculations as for $S(u)$.

\subsection{Empirical Bounds}
We empirically estimate an upper bound for $\mathbb{E}_{X\geq1}[X]\mathbb{E}_{X\geq1}\left[\frac{1}{X}\right]$ and $\mathbb{E}_{X\geq1}[X^2]\mathbb{E}_{X\geq1}\left[\frac{1}{X^2}\right]$ for a Poisson variable by varying $\lambda$. The results are presented in Figures \ref{fig:estimation1} and \ref{fig:estimation2}.
\begin{figure*}
    \centering
    \includegraphics[width=0.98\linewidth]{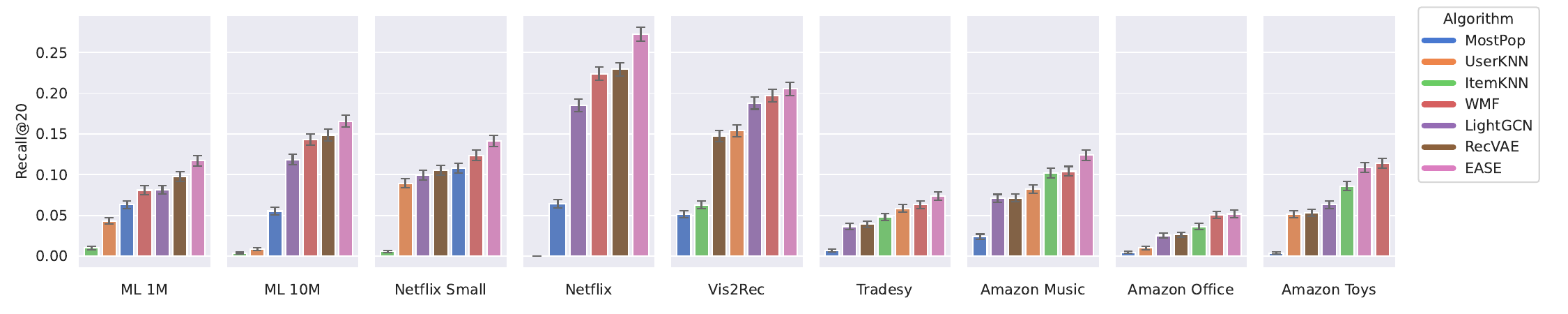}
    \caption{Overall algorithms performance across datasets, measured in \texttt{Recall@20} with confidence intervals at 95\%.}
    \label{fig:global_results}
\end{figure*}
\section{Data Processing and Training}
\subsection{Data Processing}
\label{sec:dataprocessing}
Following standard practices \cite{exploring_meng_2020}, we binarize non-binary ratings, which are all in $[1, 5]$, by setting $x_{ui} = \mathds{1}(r_{ui} > 3)$. A 5-code is extracted from the datasets, i.e. a subset of users and items with at least 5 interactions, by sequentially filtering out users and items with less than 5 interactions until convergence. This pre-processing makes the datasets more compact, leading to a size distribution described in Table \ref{tab:datasets}.
We sample 10'000 test users from each dataset, isolate their last interaction as the test set, and the second-to-last interaction as a validation set. For datasets with less than 10'000 users, all users are considered test users.
For each dataset, the Mean Surprise and Conditional Surprise are computed for each user. This allows us to easily create the segments $\mathcal{D}[\alpha, \beta]$ for multiple values of $\alpha$ and $\beta$, both in the train and test set. Since most algorithms cannot handle users that are not in the train set, we ensure that test users always appear in the train set. However, some items of the test set may not always be present in the train set, which makes the task more complicated.

We found that in most implementations of recommender data pre-processing, the data was filtered by first removing the items that were consumed by less than $k$ users, then removing the users that consumed less than $k$ items. This is a problem since with this ordering, the remaining items could be consumed less than $k$ times. 

\subsection{Optimal Hyperparameters}
The optimal hyperparameters found by \texttt{optuna} on 50 runs for each combination of dataset and algorithm, optimizing on the \texttt{Recall@20} of the validation set, are presented can be found through the anonymized repository. 
\begin{table*}[h!]
\centering
\begin{tabular}{lcccccccccccc}
\toprule
\textbf{Dataset} & \textbf{5} & \textbf{6} & \textbf{7} & \textbf{8} & \textbf{9} & \textbf{10} & \textbf{11} & \textbf{12} & \textbf{13} & \textbf{14} & \textbf{15} \\
\midrule
Netflix        & 14.36 & 17.51 & 20.88 & 24.43 & 28.18 & 32.17 & 36.36 & 40.48 & 44.80 & 49.48 & 54.03 \\
Netflix Small & 7.56  & 9.26  & 11.51 & 13.15 & 14.70 & 17.11 & 19.27 & 21.84 & 23.52 & 25.75 & 27.91 \\
ML-1m          & 7.42  & 8.24  & 9.65  & 10.96 & 12.06 & 13.08 & 14.55 & 15.69 & 16.66 & 17.82 & 19.31 \\
ML-10m         & 13.61 & 15.71 & 17.68 & 19.58 & 21.76 & 23.74 & 26.08 & 28.03 & 30.36 & 32.46 & 34.72 \\
Tradesy        & 3.27  & 3.73  & 4.29  & 5.10  & 5.84  & 6.45  & 6.89  & 7.47  & 8.52  & 9.41  & 9.81  \\
Vis2Rec        & 2.63  & 2.98  & 3.39  & 3.68  & 3.91  & 4.24  & 4.52  & 4.99  & 5.41  & 5.77  & 6.03  \\
A. Music  & 3.50  & 4.16  & 4.84  & 5.36  & 5.96  & 6.72  & 7.07  & 7.85  & 8.66  & 9.80  & 10.55 \\
A. Toys   & 2.57  & 3.41  & 4.34  & 5.32  & 6.49  & 7.68  & 8.83  & 10.26 & 11.78 & 13.42 & 15.23 \\
A. Office & 6.56  & 7.96  & 9.51  & 10.89 & 12.44 & 14.17 & 15.71 & 17.41 & 19.69 & 21.60 & 24.06 \\
\bottomrule
\end{tabular}
\caption{Signal-to-noise (SNR) values of the mean conditional surprise across different datasets, for varying profile sizes $k$. All ratios are >1, indicating that even for cold start users, the estimation of their conditional surprise is reasonable.}
\label{tab:dataset_results}
\end{table*}
\section{Results}
\label{sec:results}
\subsection{Overall Performance}

The overall results are presented in Figure \ref{fig:global_results}. All combinations of datasets and algorithms have been benchmarked except UserKNN on Netflix. Due to the amount of users, the method saturates the 1.4 TB of RAM available for experiments. 

Figure \ref{fig:global_results} shows two global trends.
First, there are substantial performance differences between movie datasets and e-commerce datasets. This can be explained by the density of their associated consumptions (see Table \ref{tab:datasets}), which is much higher for movie datasets. 
Second, performances are better when the dataset size increases: the performances of the algorithms are overall better on ML 10M and Netflix than, respectively, on  ML 1M and  Netflix Small. This is not trivial since increasing the size not only increases the number of samples (users) but also the number of items, which theoretically makes the task harder.

\texttt{EASE} provides the best performances on almost all datasets. \texttt{WMF} is second everywhere except for the biggest sets, where \texttt{RecVAE}, leverages better the amount of data at its disposal, as expected from a deep approach. However \texttt{RecVAE} and \texttt{ItemKNN} are more sensitive to the number of items and users. In particular, \texttt{RecVAE} performs better when the number of items is below the number of users. Conversely, \texttt{ItemKNN} performs better when the number of users is below the number of items.

\subsection{Signal-to-Noise Ratio (SNR) Analysis for Cold-Start Users}
\label{sec:snr_analysis}
As mentioned in the main paper, we conducted an analysis to verify that the high variance of $CS(u)$ observed for users with few interactions (i.e., cold-start users) reflects a true behavioral signal rather than a statistical artifact of noisy estimation.

\paragraph{Methodology}
We compute a Signal-to-Noise Ratio (SNR) as the ratio of inter-user signal variance to intra-user noise variance. For each profile size $k$, we define:
\begin{itemize}
    \item \textbf{Signal Variance} $\text{Var}_{\text{signal}}(k)$: The variance of $CS(u)$ scores across all users with $|u|=k$.
    \item \textbf{Noise Variance} $\overline{\text{Var}_{\text{noise}}}(k)$: The average squared standard error of the $CS(u)$ estimate for users with $|u|=k$.
\end{itemize}
The SNR is then $\text{SNR}(k) = \text{Var}_{\text{signal}}(k) / \overline{\text{Var}_{\text{noise}}}(k)$.

\paragraph{Results}
The results are conclusive: for all datasets, the SNR is substantially greater than 1, even for the sparsest users with only 5 items. For example, on the Netflix dataset, the SNR for users with 5 items is 14.36, meaning the true behavioral signal is over 14 times stronger than the measurement noise. This provides direct, quantitative evidence that our measure is a reliable tool for analyzing user coherence, even in the most challenging, data-scarce scenarios.

\subsection{Impact on Performance}
All regression are run in R, using the \texttt{glm} function, with a \texttt{binomial} law with \texttt{logit} link. The \texttt{simex} package is used to incorporate the variance of the variables, and \texttt{margins} to get the marginal effects. 

The choice of the regression to make, in particular which dependencies between variables (such as $S(u)\times CS(u)$), was motivated primarily by the model with the lowest AIC score. Consistently, the model with all the product variables met our criterion.

We also found an important aspect in modeling these logit regressions was to put a threshold on the variable $|u|$. In accordance with what we stated in the main paper for the impact of our measures on performances, this can be justified by the fact that up until a certain point, adding more items to the item set of a user can only help us to cover all their tastes. Once all the interests of a user are well represented in their item set, then we expect $|u|$ to have less importance. Indeed, thresholding $|u|$ led to models with a lower AIC but also removed the heteroscedasticity of the residuals, i.e., the dependence between the variance of the errors and the predictor variables.

\end{document}